\documentclass[a4paper,twoside]{article}

\pdfoutput=1

\usepackage[utf8x]{inputenc}
\usepackage{graphicx}
\usepackage{subcaption}
\usepackage{calc}
\usepackage{amssymb}
\usepackage{amstext}
\usepackage{amsmath}
\usepackage{amsthm}
\usepackage{multicol}
\usepackage{apalike}
\usepackage[bottom]{footmisc}

\usepackage[hyphens]{url}
\usepackage{hyperref}
\usepackage{enumitem}
\newlist{questions}{enumerate}{2}
\setlist[questions,1]{label=\textbf{RQ\arabic*.},ref=RQ\arabic*}
\setlist[questions,2]{label=(\alph*),ref=\thequestionsi(\alph*)}
\newcommand{\etal}{et\,al.\ }
\usepackage{wrapfig} 
\usepackage{xcolor, colortbl} 
\usepackage{booktabs} 

\usepackage{SCITEPRESS}     

\begin{document}

\title{Efficient aggregation of face embeddings for decentralized face recognition deployments (extended version)
	\thanks{An extended version of this paper is published at Advances in Artificial Intelligence and Machine Learning 3, 1, pp. 693--711, 2023~\cite{bib:2023-hofer-aaiml}.}
}

\author{\authorname{Philipp Hofer\sup{1}\orcidAuthor{0000-0002-7705-9938}, Michael Roland\sup{1}\orcidAuthor{0000-0003-4675-0539}, Philipp Schwarz\sup{2}\orcidAuthor{0000-0002-8364-4850} and Ren\'e Mayrhofer\sup{1}\orcidAuthor{0000-0003-1566-4646}
}
\affiliation{\sup{1} Johannes Kepler University Linz, Institute of Networks and Security, Austria}
\affiliation{\sup{2} Johannes Kepler University Linz, LIT Secure and Correct Systems Lab, Austria }
\email{\{philipp.hofer, roland, rm\}@ins.jku.at, philipp.schwarz@jku.at}
}


\abstract{ 
	Biometrics are one of the most privacy-sensitive data.
	Ubiquitous authentication systems with a focus on privacy favor decentralized approaches as they reduce potential attack vectors, both on a technical and organizational level.
	The gold standard is to let the user be in control of where their own data is stored, which consequently leads to a high variety of devices used.
	Moreover, in comparison with a centralized system, designs with higher end-user freedom often incur additional network overhead.
	Therefore, when using face recognition for biometric authentication, an \emph{efficient} way to compare faces is important in practical deployments, because it reduces both network and hardware requirements that are essential to encourage device diversity.
	This paper proposes an efficient way to aggregate embeddings used for face recognition based on an extensive analysis on different datasets and the use of different aggregation strategies.
	As part of this analysis, a new dataset has been collected, which is available for research purposes.
	Our proposed method supports the construction of massively scalable, decentralized face recognition systems with a focus on both privacy and long-term usability.
}

\onecolumn \maketitle \normalsize \setcounter{footnote}{0} \vfill

\section{Introduction}
\label{ch:introduction}

In 2021 alone, it is estimated that 14.8 zettabytes of data have been created and stored~\cite{datageneration}.
Applications processing this data do not have to disclose their existence, thus it is not possible to know about every data-processing system.
However, the public knows about at least some of these systems, as they have an immediate effect on individuals: There are many databases featuring an extensive amount of highly personal data already in production for many years, such as the Indian Aadhaar system~\cite{aadhaar}, while new ones, such as in the Moscow metro~\cite{moscow}, are being constantly rolled out.
These biometric recognition deployments are prime examples of city- or nation-scale ubiquitous systems that already bridge the digital and physical worlds through their use of (biometric and other) sensor data for deriving decisions about which physical world interactions people are authorized for.

Different systems use (or create) different kinds of data. Even though individuals can be profiled with a lot of different types of data, some data are more privacy-sensitive than others.
Especially due to the trend of increasing quantity and quality of such systems, it is especially critical to think about systems that contain highly sensitive personal data ---
even more so, as these systems may depend on public acceptance.
One of the most intrusive types of data are biometric features, as they hardly ever change.
If a system that contains, for example, fingerprints or faces, is compromised, people cannot switch to new fingerprints or faces as would be best practice for password breaches. 

In general, data can be stored in a central location or in a decentralized manner, i.e.\ spread across multiple locations.
Unfortunately for individuals' privacy, the backbone for most current systems is a central database with highly sensitive personal data, such as biometrics (including the Aadhaar system as one of the largest collections of biometric data). 
This makes such systems especially vulnerable to multiple attack vectors:
a) first and foremost, people have to trust the provider.
Since all data is stored and every request appears at this single point, the provider could potentially determine sensitive (behavioral) patterns of individuals and thus massively invade their privacy; and
b) a central place with possibly billions of records of users represents an excellent target for (technical, legal, or organizational --- including insider) attack.
Even under highest security constraints, data breaches are permanently happening, even (or especially) with the largest entities\footnote{The biggest known data breaches are visualized at \href{https://informationisbeautiful.net/visualizations/worlds-biggest-data-breaches-hacks/}{https://informationisbeautiful.net/visualizations/worlds-biggest-data-breaches-hacks}.}.

In order to reduce this massive attack surface of having everything stored in a single place, we instead propose to design decentralized systems.
Instead of having a single point with all information, the individual should ideally be allowed to choose who is managing their (personal) data.
If a person trusts, e.g., their government or bank, it is reasonable for them to also allow these entities to store and manage their digital identity, including their biometric templates.
On the other hand, the individual should have the chance to manage their data on their own devices and to move from one service provider to another (in line with the right to data portability under GDPR~\cite{gdpr}). 
The specific trade-offs of using different service providers or own devices for hosting and managing personal identity data is highly context dependent and should therefore always be under control of each individual user.

However, these kinds of distributed systems suffer from different drawbacks.
The integrity of each device that manages an individual must be verified to prevent identity theft by malicious devices.
One of the main points of creating a decentralized system is that the user should be in charge of deciding where their personal data is stored. 
Thus, the user should be able to pick the most trustworthy provider. 
Stressing (hardware) requirements might favor fewer, big players who have the necessary resources, and therefore systems with fewer requirements will support a larger provider distribution.
Since biometrics are one of the most privacy-sensitive data points and efficiency is crucial for the successful deployment of large-scale systems, this paper focuses on the computational complexity required for biometric authentication systems.

In many instances, data collection is not the problem, but
processing large quantities of data proves challenging.
In the case of face recognition, it is likely that there are many images available for a specific individual, for example obtained through mass-scale web scraping.
State-of-the-art face recognition pipelines typically start by extracting embeddings that represent feature vectors of individual faces through the use of machine learning models (currently often based on deep neural networks); 
subsequently, they can calculate the similarity to all of these images by again deriving embeddings and comparing them to previously stored templates and each other.
This is inefficient on two different levels: First, decentralized systems strive for as little computational requirements as possible.
Therefore, requiring multiple (on a holistic system point of view, redundant) similarity calculations would hurt both provider diversification and hinder small providers in serving larger quantities of users.
Ideally, one could combine different aspects of these multiple embeddings extracted from face images with as little data as possible.
Since research on still image face recognition is quite extensive, and an embedded camera sensor device can often derive embeddings of the currently visible person on-line, creating a new, aggregated embedding based on all images available of an individual would not change the backbone of state-of-the-art face recognition pipelines.
Secondly, having a single (aggregated) embedding for an individual, and thus not depending on multiple similarity computations minimizes network traffic, as only a single embedding needs to be sent across the network, which is especially significant for decentralized, embedded systems.
Additionally, we argue that aggregating face embeddings as stored by recognition systems can also be a privacy advantage: 
while embeddings of single face images can potentially be reversed to compute (an approximation of) the original input image\cite{fabian2020anonymizing}, reversing the aggregate of multiple different embeddings would necessarily create an artificial face image that overlays aspects of all the aggregated input. 
While this may still represent the person sufficiently well for further correlation analysis, it should mask the original inputs that could otherwise be used for more direct matching. 
However, a detailed, quantitative analysis of this privacy aspect is outside the scope of this and subject to future research.

In this paper, our focus is on evaluating different methods of aggregating face embeddings (Section~\ref{sec:diff-setting}) from an efficiency and accuracy point of view.
Furthermore, we test the limit of sufficient image quantity and analyze whether there is a clear point where adding additional images does not significantly increase face recognition accuracy (Section~\ref{ssec:limit}).
Last but not least, in order to verify if using multiple images in different settings boosts accuracy significantly, we propose a new in-the-wild dataset, where subjects take around 50 images of themselves in a single setting, which only takes around 3 seconds to achieve practical usability (Section~\ref{sec:dataset}). 
Additional images in radically different settings are used as approximation of the true embedding to verify the performance improvements (Section~\ref{sec:ssp}).

\section{Multiple images in face recognition pipelines} 

In order to evaluate and compare different face recognition methods, they are tested against public datasets.
Many of these face recognition datasets typically have two properties:
\begin{enumerate}
	\item High quality: As the datasets are created with training face recognition models in mind, the images of a person mainly consists of a portrait image in a fairly high resolution.
	\item High quantity of people: Typically, bias of neural networks becomes less the more images are used.
	Therefore, datasets strive for a high amount of images.
\end{enumerate}

Most datasets define a fixed set of pairs of images to allow for objective evaluation of face recognition methods.
With this strategy, a single image is used as template in state-of-the-art face recognition pipelines.
This template is then compared with positive (same person) and negative (different person) matches.
This approach tests one important metric of face recognition: How well it is performing on still images.
Compared to more complex scenarios, only testing on still images is efficient at runtime, which decreases computation time to evaluate the accuracy on a dataset.
However, there are different aspects this method does not test, such as how to handle multiple images or even video streams of a person.

In reality, these ignored aspects are essential, as live-images from cameras do not produce high-quality images similar to the images from many available and commonly used face recognition datasets.
Instead, the person-camera angle is far from optimal, the person is not directly in front of the camera and thus the face is quite small.
Furthermore, the face can be occluded, e.g. with a scarf, sunglasses, or hair.
In these real-world settings, face recognition pipelines have a harder time recognizing people than with public datasets, although new datasets try to represent these challenges.
Nevertheless, there is a potential benefit of real-world scenarios: There are many images of a single person available, as the person is presumably visible for (at least) many seconds and thus a camera is able to capture significantly more than one image.

One way of bridging the gap of having multiple images of the same person and being of lower quality is to merge the embeddings obtained from multiple images into a single embedding.
More accurate templates, by definition, lead to accuracy improvements of face recognition.
The idea behind using multiple images is that it is not possible to capture a perfect representation of a face in a single picture due to various reasons:
\begin{itemize}
	\item The image only captures part of the face.
	It is technically not possible to cover frontal and profile pictures in a single 2D image.
	\item External conditions, such as lighting, camera quality, and insolation, change.
	\item People themselves change over time: Growing a beard, getting wrinkles or a new hair cut, putting on makeup, getting pierced or tattooed, getting a scar, or having surgical interventions.
	\item Different accessories, such as (sun) glasses, headgear, earrings, or masks are worn.
\end{itemize}

While a single image cannot account for all these different settings, multiple images can capture different face areas and settings.
Therefore, using multiple images provides more information about the individual's face, and we therefore expect an increased accuracy.
As introduced in Section~\ref{ch:introduction}, due to hardware and network constraints, comparing the current live-image with multiple embeddings of the same person is not favorable in many situations.
For an efficient face recognition pipeline, it would be best to only have a single embedding which is used as template for a person. 
This would allow the system to make use of the vast literature on single face image recognition.

In contrast to this single-embedding approach, in recent years other work is published in the domain of video face recognition~\cite{rao2017attention,rivero2021ordered,zheng2020automatic,liu2019feature,gong2019video}.
Most of these papers propose an additional neural network to perform the weighting of different embeddings~\cite{liu2019feature,rivero2021ordered,yang2017neural,gong2019video}.
Especially on embedded devices, these additional networks have a significant runtime impact, as they need to perform an additional inference step.
In order to be runtime-efficient even on embedded hardware, this paper focuses on creating a single embedding.

In state-of-the-art face recognition tools, embeddings are high-dimensional vectors.
If multiple embeddings should be aggregated to a single one, this opens up the questions:

\begin{questions}[start=1]
	\item  How do we (numerically) \emph{best} aggregate the embeddings, and is this aggregation actually increasing face recognition performance? How can we define \emph{best}?\label{rq:numagg}
\end{questions}

\begin{questions}[start=2]
	\item After knowing how to aggregate embeddings, how many images are necessary and useful?
	Is there a point from which adding additional embeddings do not significantly increase accuracy?\label{rq:quant}
\end{questions}

\noindent
Depending on the application, there may or may not be a lot of data available for each person. Therefore, in many situations (e.g. enrollment of a user) it could make the process significantly easier if the data can be recorded in a single session and therefore featuring only one setting.
This leads to the question:
\begin{questions}[start=3]
	\item Is it beneficial to use different settings?
	Is it worth creating images with and without (typical) accessories, such as face masks, glasses, and scarfs?
	\label{rq:diff-settings}
\end{questions}

\noindent
Similarly, it might be unrealistic to expect many images in various settings from a new user.
Having to verify that these different settings really belong to the same person makes it even more complicated. 
It is easy and practical to capture a couple of images in one place.
\begin{questions}[start=4]
	\item Is it enough if we use only images while we rotate our heads for the aggregated embedding, similar to the process of how some smartphone enroll users' faces?
	Is the accuracy increased if we include totally different settings in the aggregated embedding?
	\label{rq:dataset}
\end{questions}

\section{Embedding aggregation}
\label{sec:diff-setting}

This paper evaluates different aggregation strategies and proposes efficient ways of aggregating embeddings in order to create a single, efficient template-embedding containing as much information as possible.
If multiple images of a person are used, the position of the person of interest has to be extracted in each frame.
These positions could be either fed to a neural network which expects multiple images (or a video) as input (\textit{video-based face recognition}) or the embedding could be extracted for each frame and then aggregated (\textit{imageset-based face recognition}).

Video face recognition networks have to perform all necessary steps in a single network.
Extracting the embeddings frame-by-frame and only then aggregating these embeddings to a single template allows for a much more modular pipeline.
This goes hand-in-hand with traditional face recognition pipeline approaches, which can be separated into face detection, face tracking, and face recognition, and therefore also allow for being able to individually optimize each part.
Additionally, systems using this approach can use its vast literature, as the field of (still) image face recognition is much more advanced than video face recognition.
Therefore, in this paper, we focus on the modular approach of extracting embeddings from every image and then aggregating them.

Since we want to aggregate multiple embeddings extracted from single frames, we need an \textit{aggregation strategy}.
Literature typically calculates the mean of each dimension of the embedding, e.g. as proposed by Deng \etal~\cite{deng2019arcface}:

\[\begin{pmatrix}
	a_1\\
	a_2\\
	\vdots \\
	a_n
\end{pmatrix},
\begin{pmatrix}
	b_1\\
	b_2\\
	\vdots \\
	b_n
\end{pmatrix} \rightarrow
\begin{pmatrix}
	\text{mean}(a_1,b_1)\\
	\text{mean}(a_2,b_2)\\
	\vdots \\
	\text{mean}(a_n,b_n)
\end{pmatrix}\]

\begin{figure}[b]
	\vspace{-0.2cm}
	\centering
	\includegraphics[width=\columnwidth]{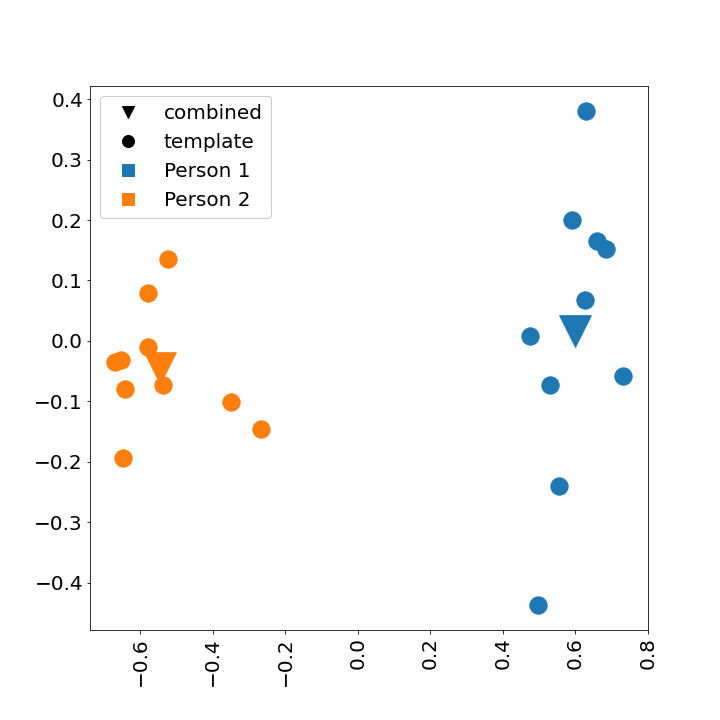}
	\caption{PCR-reduced visualization of the first two people with 30 images in the CelebA dataset. The triangle represents the calculated mean of each embedding.}
	\label{fig:instance-space-mean}
	\vspace{-0.1cm}
\end{figure}

Fig.~\ref{fig:instance-space-mean} shows the instance space of two people.
The x- and y-axis represent the PCR-reduced form of their embeddings.
The triangle represent the average of each dimension of the embedding for each person.
There is no analysis on whether it is useful to use the mean of each dimension, or if there are better approaches to aggregate embeddings.
There is not even an analysis if calculating the mean of the embeddings improve the accuracy of face recognition pipelines.
In order to verify that a mean calculation of every dimension of the embeddings increases accuracy, a baseline is needed to compare the performance of aggregated embeddings to.
In this paper, we use pre-trained state-of-the-art face-detection (RetinaFace~\cite{deng2020retinaface}) and -recognition (Arcface~\cite{deng2019arcface}) models.
Arcface receives a single image as input and creates a 512-dimensional vector.
Even though we did not explicitly test different architectures, we expect similar results on semantically similar networks.
Further work will focus on extending these experiments to multiple face detection and -recognition models.

\subsection{Dataset}
\label{dataset}
We chose to evaluate the face recognition models on the CelebA dataset~\cite{liu2015deep} which contains multiple images of thousands of people.

\begin{figure}[b]
	\vspace{-0.2cm}
	\centering
	\includegraphics[width=\columnwidth]{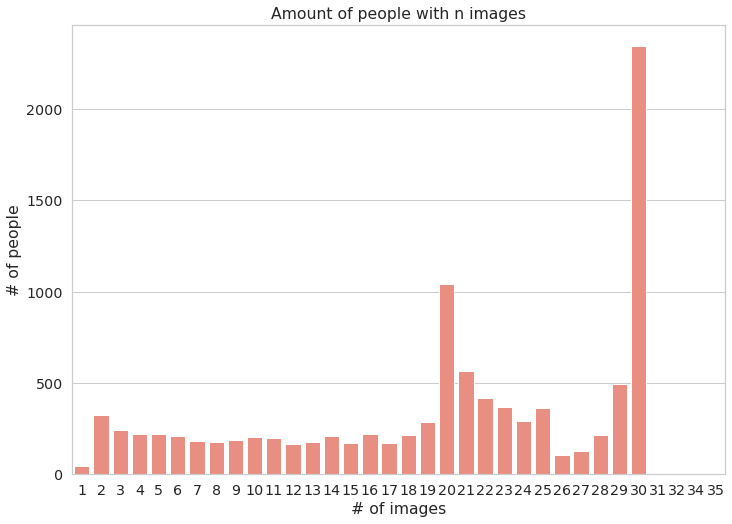}
	\caption{Distribution of number of people with respect to their number of images.}
	\label{fig:celeba-distribution}
	\vspace{-0.1cm}
\end{figure}

More specifically, the dataset contains 10.177 people.
2.343 people have exactly 30 images (Figure~\ref{fig:celeba-distribution}).

\begin{figure}[b]
	\vspace{-0.2cm}
	\centering
	\includegraphics[width=\columnwidth]{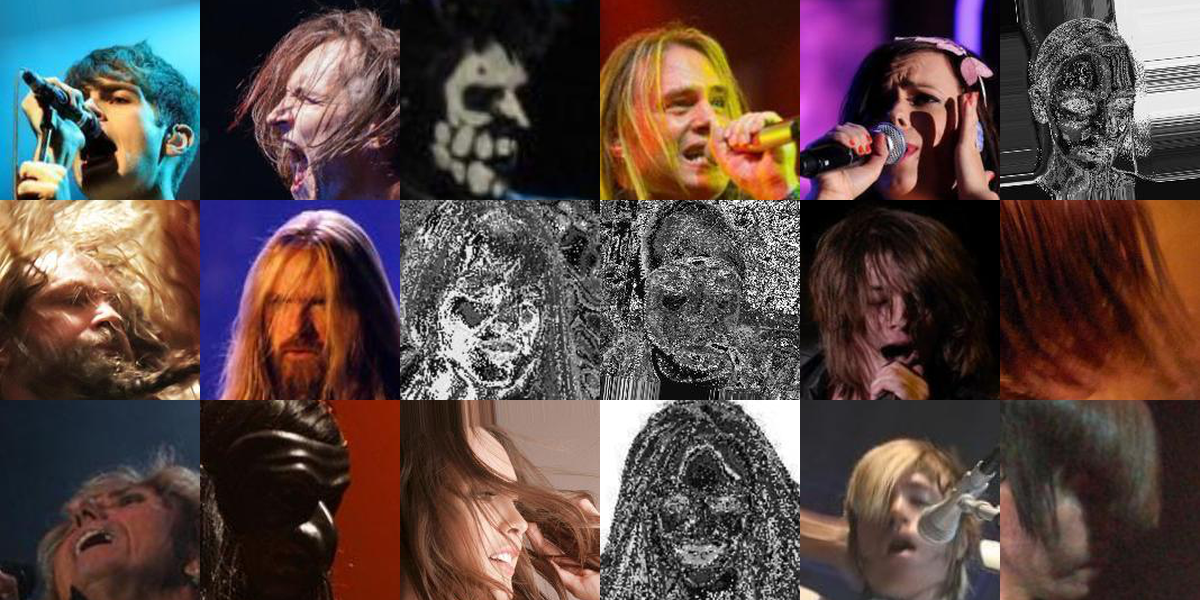}
	\caption{Example images where RetinaFace could not detect a person.}
	\label{fig:face-det-failed}
	\vspace{-0.1cm}
\end{figure}

In order to reduce the chance of having outliers, we remove all images with less than 30 images.
For consistency, we also remove people with more than 30 images.
Furthermore, we cleaned the dataset by performing face detection with RetinaFace.
From the 2.343 people with 30 images, there are 20 people which contain an image where face detection could not detect a face---mainly due to too much occlusion. 
18 randomly chosen images, where face detection did not work, are shown in Fig.~\ref{fig:face-det-failed}.
In order to have a dataset as consistent as possible, we removed all images of these 20 people, resulting in a final set of 2.323 people.

\begin{figure}[b]
	\vspace{-0.2cm}
	\centering
	\includegraphics[width=\columnwidth]{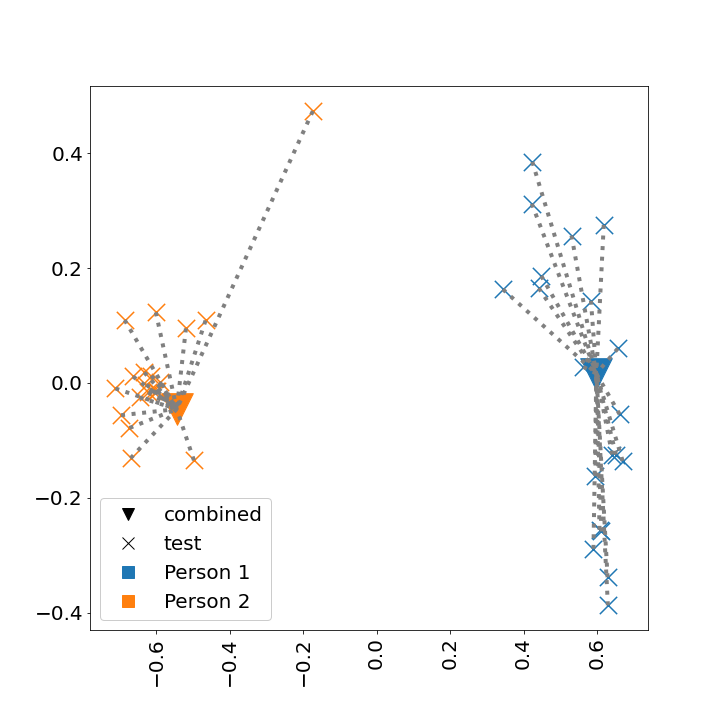}
	\caption{The sum of the dashed lines represent the error from the template (triangle) to the test images.}
	\vspace{-0.1cm}
\end{figure}

The other corner case is when multiple faces are detected in a single image.
Since the CelebA dataset has been pre-processed such that the main person is in the center of the image, we simply take the most central person and ignore the remaining ones.

In total, there are 30 images of 2.323 people each in our cleaned dataset, resulting in a total number of 69.690 images.
In order to objectively evaluate the difference between different aggregation strategies, we reserve 10 random images of each person as potential template images.
Since the dataset does not have a specific order, without loss of generality and for reproducibility, we reserve the first 10 images as potential template images.

In the first setting (\textit{baseline}), we (only) use the embedding of the first image and ignore images 2 -- 9:
\begin{equation}
	template_{baseline}(person) = emb_{person}[0].
\end{equation}
\label{eq:template_baseline}

Calculating the mean of different embeddings is only one possible strategy to aggregate embeddings.
For different methods, such as taking the minimum of each dimension, we numerically aggregate each dimension of the embeddings ($e$) from images 1 -- 9, using its respective aggregation strategy ($as$):
\begin{equation}
	 \begin{aligned}
		\begin{pmatrix}
			e_{11}\\
			e_{12}\\
			\vdots \\
			e_{1m}
		\end{pmatrix},
		\dots ,
		\begin{pmatrix}
			e_{n1}\\
			e_{n2}\\
			\vdots \\
			e_{nm}
		\end{pmatrix} \rightarrow 
		\begin{pmatrix}
			\text{as}(e_{11}, \dots, e_{n1})\\
			\text{as}(e_{12}, \dots, e_{n2})\\
			\vdots \\
			\text{as}(e_{1m}, \dots, e_{nm})
		\end{pmatrix}.
	\end{aligned}
\end{equation}
\label{eq:aggregation_strategies}

Typically, face recognition models are trained such that the L2-distance between two faces represents their (non-)similarity:

\begin{equation}
	dist(emb_1, emb_2) = \sqrt{\sum_{i=1}^n (emb_1-emb_2)^2}.
\end{equation}
\label{eq:dist}

Applications set a threshold for their specific task, under which two faces are recognized as the same person.
This decision is based on their safety requirements.
For security-critical applications the threshold should be lowered, which results in fewer false-positives (but potentially more false-negatives).

\begin{equation}
	\begin{aligned}
	isSamePerson(emb_1, emb_2) =  \\
	\begin{cases}
		1, & \text{if } dist(emb_1, emb_2)\leq \textbf{threshold} \\
		0, & \text{otherwise}.
	\end{cases}
	\end{aligned}
\end{equation} 

In order to compare the aggregation strategies, we take the average distance of the template to each embedding from images 11 -- 30 as our metric:

\begin{equation}
		error = \frac{\sum_{p \in people} \sum_{emb_{test} \in testEmbs} dist(emb_{test}, p_{template})}{len(people) \times len(testEmbs)}
\end{equation}

A smaller \textit{error} represents a higher confidence of the network, that the template belongs to the test images. Semantically, the error specifies the average distance (Eq.~\ref{eq:dist}) of the template to each test image.

\begin{figure}
	\vspace{-0.2cm}
	\centering
	\includegraphics[width=\columnwidth]{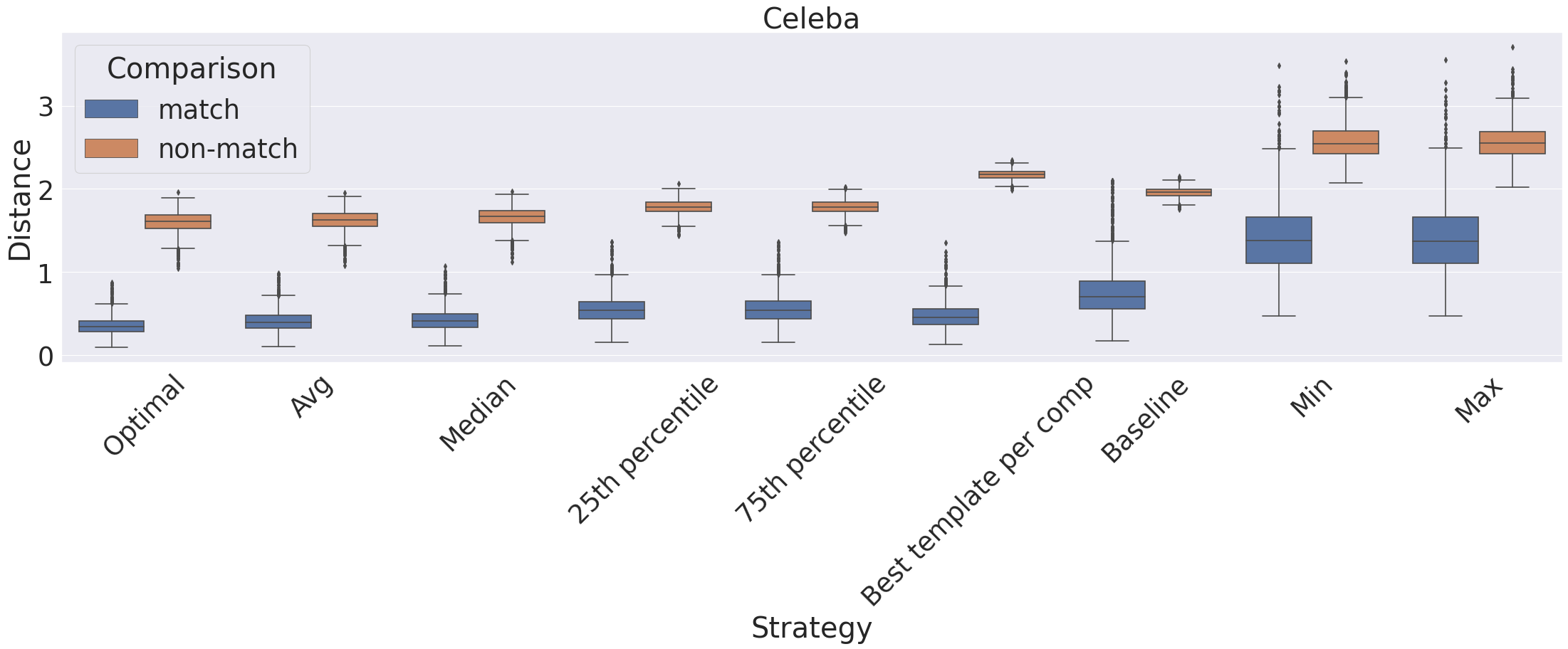}
	\caption{Average distance of template- to test-embeddings in CelebA dataset.}
	\label{fig:dist-celeba-boxplot}
	\vspace{-0.1cm}
\end{figure}

\begin{figure}[b]
	\vspace{-0.2cm}
	\centering
	\begin{subfigure}[b]{0.45\columnwidth}
		\centering
		\includegraphics[width=\columnwidth]{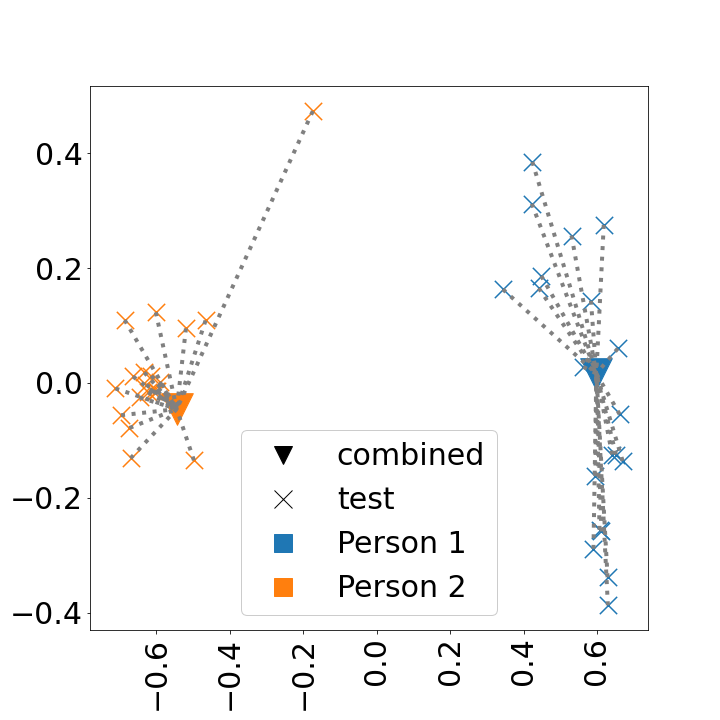}
	\caption{... the calculated mean-template.}
	\end{subfigure}
	\begin{subfigure}[b]{0.45\columnwidth}
	\centering
	\includegraphics[width=\columnwidth]{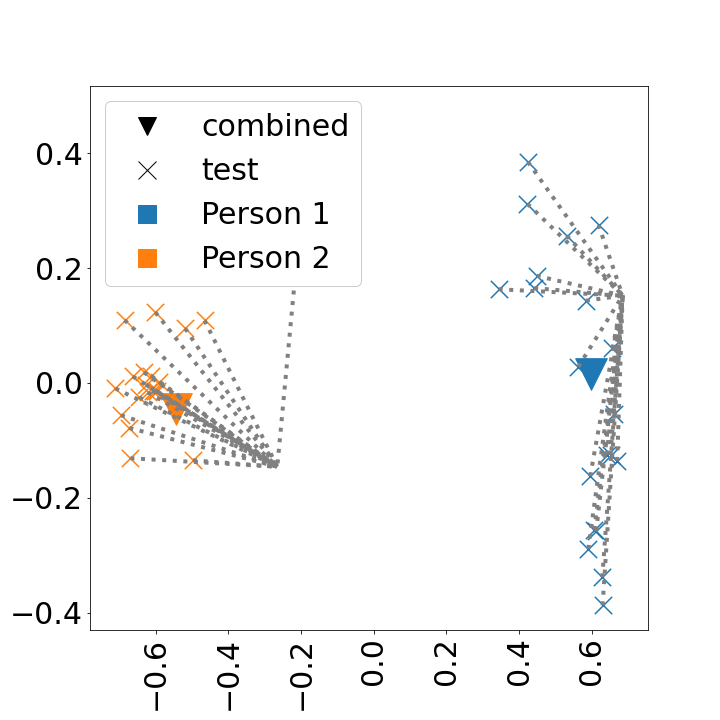}
	\caption{... the first image (baseline).}
	\end{subfigure}
	\caption{Instance space: Distance of the embeddings to ...}
	\label{fig:instance-4}
	\vspace{-0.1cm}
\end{figure}

The \textit{CelebA} row in Table~\ref{tbl:main-result} shows the resulting distance between the template- and test-embeddings.
Fig.~\ref{fig:dist-celeba-boxplot} visually represents the CelebA column.
With respect to our \ref{rq:numagg}: Except for the (cheating) \textit{optimal} setting (which we will discuss later), the best aggregation strategy is using the mean of every dimension of the embedding.
As the distance compared to the baseline is significantly lower in the \textit{average} (and \textit{median}) setting, this clearly shows the effective impact of using multiple (in our case 10) images as templates.
Aggregating multiple embeddings using the mean significantly outperforms the baseline.
Fig.~\ref{fig:instance-4} shows the intuition behind this behavior on the first two people in the dataset.
If more than a single image of a person is used, the resulting embedding approximates the optimal embedding more accurately.
The optimal embedding of a person with respect to the current test images, is the average of these test embeddings, since this would minimize the respective distance.

\begin{figure}[b]
	\vspace{-0.2cm}
	\centering
	\includegraphics[width=\columnwidth]{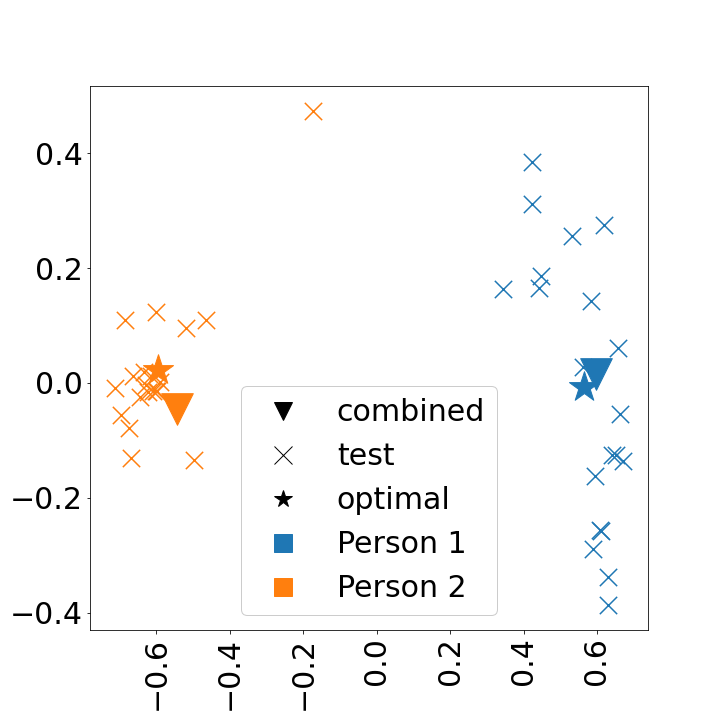}
	\caption{Optimal vs average} 
	\vspace{-0.1cm}
\end{figure}

\begin{table*}[]
	\centering
	\begin{tabular}{ll|l|l|l}
		&      & CelebA & Panshot-Normal & Panshot-Smile\\
		\hline
		Baseline        & Match& 0.748  (1.0x)& 0.617  (1.0x)& 0.612  (1.0x)\\
		& Non-Match & 1.958  (1.0x)& 1.888  (1.0x)& 1.840  (1.0x)\\
		\hline
		
		Avg        & Match& 0.410  (1.8x)& 0.404  (1.5x)& 0.425  (1.4x)\\
		& Non-Match & 1.622  (1.2x)& 1.520  (1.2x)& 1.476  (1.2x)\\
		\hline
		Median        & Match& 0.422  (1.8x)& 0.423  (1.5x)& 0.443  (1.4x)\\
		& Non-Match & 1.662  (1.2x)& 1.611  (1.2x)& 1.556  (1.2x)\\
		\hline
		Min        & Match& 1.414  (0.5x)& 2.208  (0.3x)& 2.160  (0.3x)\\
		& Non-Match & 2.567  (0.8x)& 3.107  (0.6x)& 3.006  (0.6x)\\
		\hline
		Max        & Match& 1.409  (0.5x)& 2.186  (0.3x)& 2.166  (0.3x)\\
		& Non-Match & 2.564  (0.8x)& 3.080  (0.6x)& 3.010  (0.6x)\\
		\hline
		25th percentile        & Match& 0.552  (1.4x)& 0.557  (1.1x)& 0.574  (1.1x)\\
		& Non-Match & 1.781  (1.1x)& 1.718  (1.1x)& 1.661  (1.1x)\\
		\hline
		75th percentile        & Match& 0.552  (1.4x)& 0.560  (1.1x)& 0.581  (1.1x)\\
		& Non-Match & 1.781  (1.1x)& 1.716  (1.1x)& 1.668  (1.1x)\\
		\hline
		\rowcolor{gray!30}
		Optimal        & Match& 0.354  (2.1x)& 0.383  (1.6x)& 0.382  (1.6x)\\
		\rowcolor{gray!30}
		& Non-Match & 1.604  (1.2x)& 1.501  (1.3x)& 1.443  (1.3x)\\
		\hline
		\rowcolor{gray!30}
		Best template per comp        & Match& 0.471  (1.6x)& 0.170  (3.6x)& 0.214  (2.9x)\\
		\rowcolor{gray!30}
		& Non-Match & 2.173  (0.9x)& 2.195  (0.9x)& 2.121  (0.9x)\\
	\end{tabular}
	\caption{This table shows the average distances of the template embedding to the test embeddings with respect to different aggregation strategies. The value in brackets represents the factor of the distance of that strategy compared to the baseline. A factor of 2 means that the average distance of the baseline is twice as high as this particular aggregation strategy. For \textit{match}es, a higher factor is favorable, while for \textit{non-match}es, a lower factor is better. The gray rows are displayed for comparison reasons only, as they cheat and use information not available in production.}
	\label{tbl:main-result}
\end{table*}

\begin{figure*}
	\vspace{-0.2cm}
	\centering
	\includegraphics[width=\columnwidth]{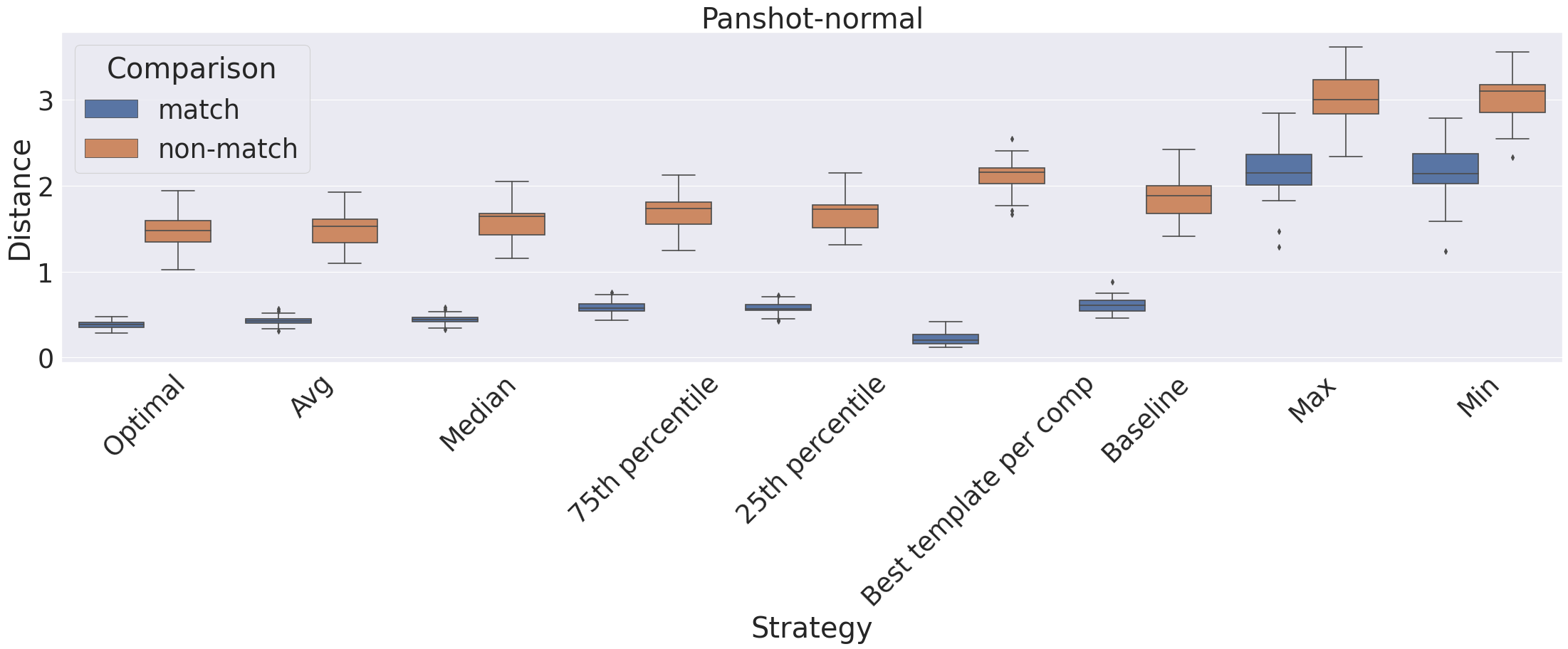}
	\caption{Average distance of template- to test-embeddings in Panshot-Normal dataset.}
	\label{panshot-normal}
	\vspace{-0.1cm}
\end{figure*}

\subsection{Image quantity plateau}
\label{ssec:limit}

In the previous section, it was shown that using the average of 10 images significantly outperforms a single image used as template.
Naturally, the question arises if the accuracy still increases if more images are used.
Is there a limit above which additional images will not further improve accuracy (\ref{rq:quant})?

To answer this question, we need a dataset with more images of the same person.
For this purpose we used the LFW dataset~\cite{huang2008labeled} as it contains hundreds of images of the same people.
In particular, we use the 5 people in the LFW dataset who have more than 100 images.

\begin{figure}[]
	\begin{subfigure}{.45\columnwidth}
		\includegraphics[width=\textwidth]{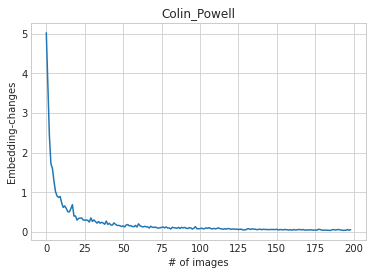}
		\centering
	\end{subfigure}
	\begin{subfigure}{.45\columnwidth}
		\includegraphics[width=\textwidth]{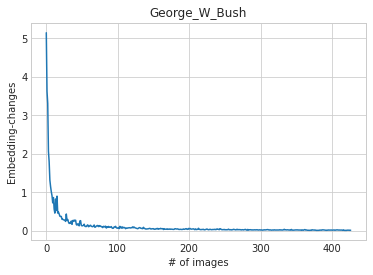}
		\centering
	\end{subfigure}
	
	\caption{Numeric embedding differences shown for 2 people from the LFW dataset.}
	\label{fig:amount_embs}
	\centering
\end{figure}

For each person, the embedding of the first image serves as the starting point.
Next, the embedding of the second image is extracted.
The first point of each plot in Fig.~\ref{fig:amount_embs} represents the sum of the difference between these two embeddings.
We then combine all previously used embeddings into our template.
Then, we extract the embedding of the next image, calculate its difference to the template, and plot the value.
We continue with this approach until we used every available image.

Interestingly, it looks like a (fuzzy) inverse log function.
Intuitively, this makes sense as new images contain a lot of new information in the beginning, but after the template consists of many aggregated images, a new image cannot provide as much new information as in the beginning.
Furthermore, there seems to be a limit of roughly 50 images, after which the embedding is not changing significantly anymore.
Another aspect to point out, is that there are some \textit{upticks} in the graph.
After looking at the specific images which cause these effects, they all present a new variation of the face (either a new face-angle or different accessories).

\subsection{Face-image requirements} 
\label{sec:diff-angle}

Section~\ref{sec:diff-setting} used images of the same person in different settings, such as different hairstyles, lighting, and location.
For the most part, the dataset consists of frontal images as the person is directly looking into the camera.
Some modern smartphones provide the ability to unlock the phone by rotating the phone around the head.
This is probably not only used to detect the liveness of the person, but also to increase the amount of information gained from the camera.
Is the difference in angle from this type of recording enough to utilize the benefit of combining embeddings discussed so far (\ref{rq:diff-settings})?

Therefore, we did a similar analysis on a different dataset: Pan Shot Face Database (PSFD)~\cite{findling2013towards}.
This dataset features 30 participants from 9 perspectives.
Every perspective contains 5 \textit{look directions} (straight, slightly top left, slightly top right, slightly bottom right, and slightly bottom left) and 4 distinct \textit{facial expressions} (normal, smiling, eyes closed, and mouth slightly opened).
This gives us 5,400 images to work with.

For the first test, we used all images with a \textit{normal} face expression as template and evaluated its average distance to all other images.
The result is visible in the \textit{Panshot-Normal} row in Table~\ref{tbl:main-result}.

People in this dataset seem to be easier recognized compared to the CelebA dataset, which is reflected in a lower average distance (Table~\ref{tbl:main-result}).
For the CelebA dataset, the template which consists of 10 images is performing 1.8 times better than if only a single image is used as template.
Interestingly, on our new dataset this improvement is in the same order of magnitude: 1.5 times better.

In order to simulate real-world templates, images are professional portrait photographs (e.g. used as profile images) of the subject.
In the second scenario templates are created with images of the smiling person.
The outcome of this \textit{Panshot-Smile} setting, is not significantly different to the original \textit{Panshot-Normal} setting (c.f.~Table~\ref{tbl:main-result}).
Thus, it does not make a significant difference which facial expression the person put on while creating template images.

\section{In-The-Wild Face Angle Dataset}
\label{sec:dataset}

In our experiments so far, we used images of the same people in different settings, as these are the most common images provided by available datasets.
In practice, however, it would seem convenient for both the provider and the individual to only use images taken at the time of physical enrollment.
The provider would benefit by ensuring that the individual is not spoofing the system, e.g., by using images from other people\footnote{
	Note that providing wrong or even specifically manufactured images to the enrollment process could have multiple goals: 
	in addition to the obvious enrollment of a set of images containing faces of two or multiple persons to make them all recognized as a single system user, malicious users might try to attack the embedding computation or matching approaches directly by exploiting model weaknesses through specifically tampered input. 
	The exact attack vector is outside the scope of this paper, as our proposed collection of multiple images in a single, controlled enrollment session is assumed to prevent both targets at once.} ---
which would break security guarantees for all kinds of authentication systems, both with publicly issued credentials such as passports and with accounts enrolled with only a single (e.g., building access control) system.
The advantage for the user is better usability, as they do not have to provide any additional data besides their participation in the enrollment procedure.
With enrollment interaction limited to a few seconds, we argue that creating a more diverse set of input face images for improving recognition accuracy as proposed in this paper takes less effort than creating a traditional user account with setting a new password.

Unfortunately, there are no publicly available datasets that systematically contain both images of people in the same setting (e.g., only rotating the head, as performed for some mobile phone face authentication implementations) and also images in different settings.
In order to test our hypothesis of only using a single setting while additional images of the same person in different settings do not increase accuracy, we created a new dataset, which we called \textit{In-The-Wild Face Angle Dataset}. 
We will also use this dataset to answer~\ref{rq:dataset}.
Inspired by Datasheets for Datasets~\cite{gebru2021datasheets}, we describe the dataset on our website: \href{https://digidow.eu/experiments/face-angle-dataset}{digidow.eu/experiments/face-angle-dataset}.

\section{Single setting performance}
\label{sec:ssp}

In order to test the increased performance if multiple images recorded in a single session are used, we calculated a rolling average of the template images.
The first data point for each person is equal to the first embedding.
The second data point is the average of the first two embeddings.
The last data point is the average of all embeddings of this particular person.

In order to quantify the performance, the 10 images of each person in different settings are not used as template images.
Instead, the average distance between the rolling average of the template images and test images is calculated.

\begin{figure}[]
\centering
\begin{subfigure}[b]{0.45\textwidth}
	\includegraphics[width=\textwidth]{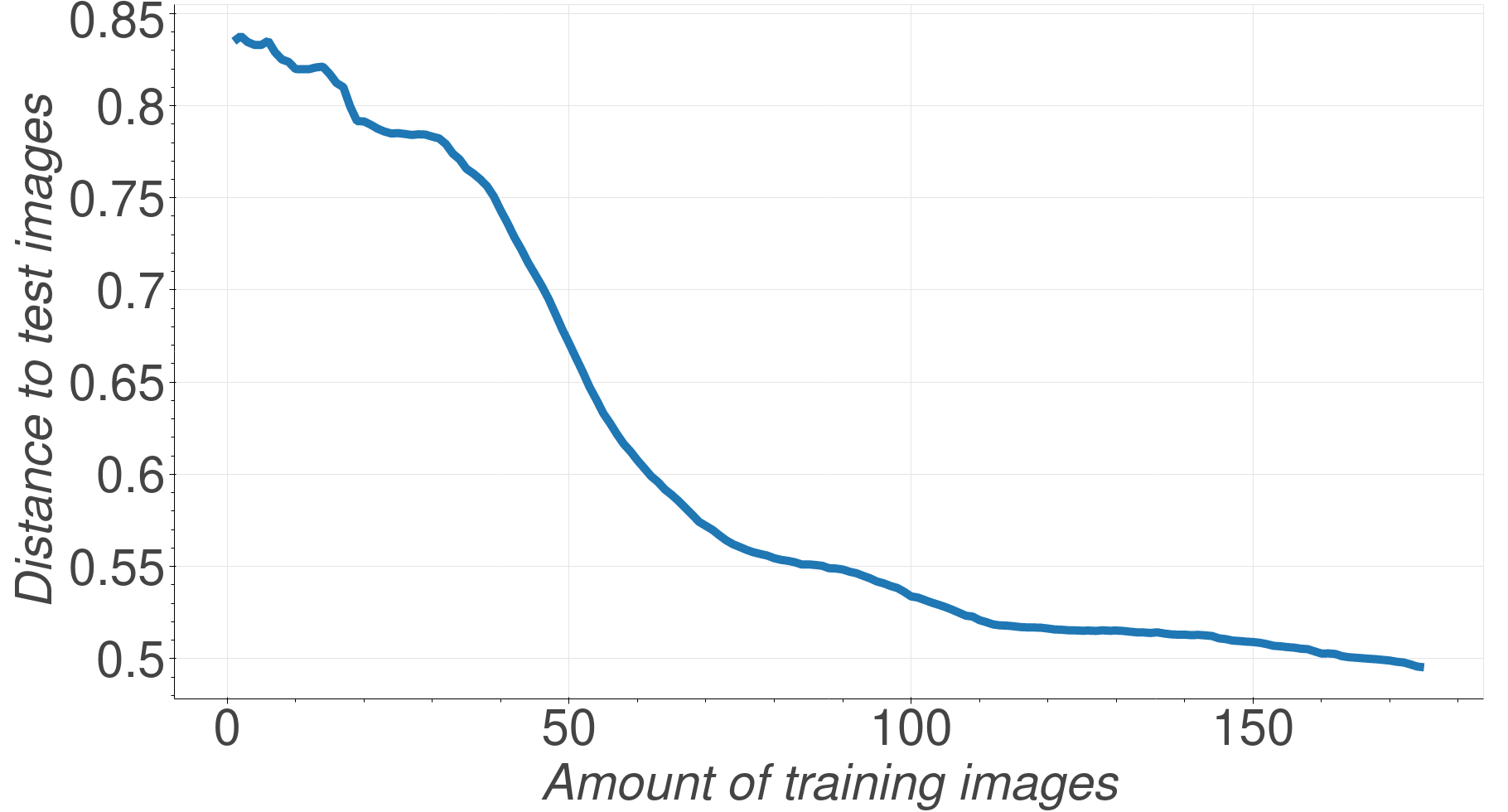}
	\caption{Template images used in sequence.}
	\label{fig:fad1}
	\centering
\end{subfigure}\vspace{5pt}
\begin{subfigure}[b]{0.45\textwidth}
	\includegraphics[width=\textwidth]{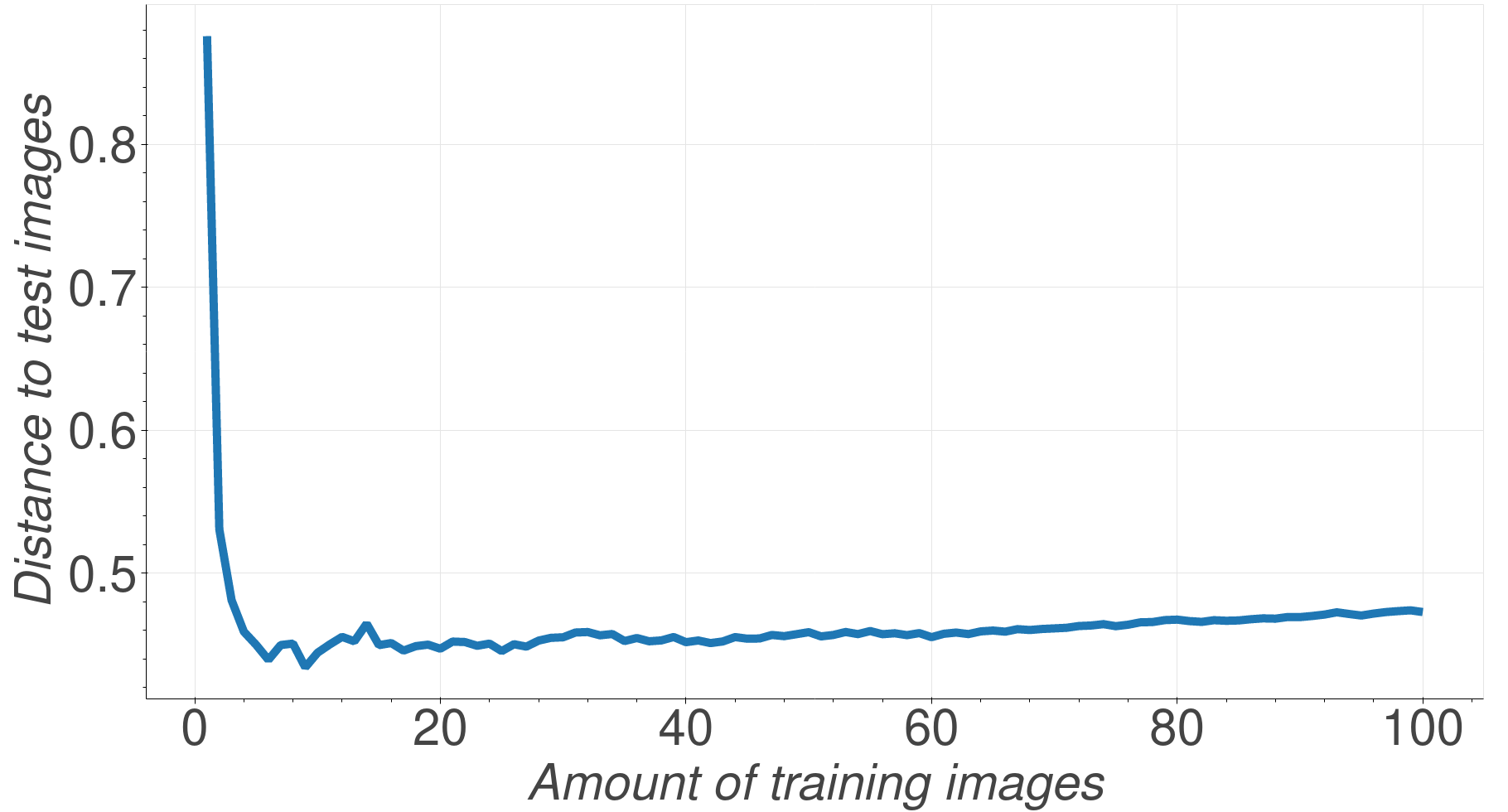}
	\caption{Greedy search for selecting the best template on every iteration.}
	\label{fig:fad2}
	\centering
\end{subfigure}
\caption{These plots show the rolling distance average of the aggregated embedding to the test images. The x-axis represents the rolling average of the embeddings from image $1$ to $x_i$. The y-axis shows the average distance to the test images.}
\label{fig:fad}
\end{figure}

The average distance between the first image of every person and their test images is, on average, 0.699.
If we not only use the first image, but rather the average of all template images, the average distance drops significantly to 0.291.
Fig.~\ref{fig:fad1} shows the average plot of a person.

Interestingly, this opposes the findings of Section~\ref{ssec:limit} as the distance grows smaller even after the limit of roughly 50 images.
We argue, that this is due to the fact that not the amount of images, but the amount of \textbf{semantically different} images are important.

To verify this, we ran the same experiment, but used only every $n^{th}$ image as template image ($n={10,20,50}$). 
The results are shown in Table~\ref{tbl:fad-nth}: The distance decreases if more images are used.
However, distance improvements are certainly not linear and are leveling off at some point.
Since the improvement from using just a few images (\textit{every 50th image}) is only marginally better than using hundreds of images (\textit{all images}) suggests, that the amount of images only plays a minor role.

\begin{table}[]
\centering
\begin{tabular}{@{}l|l@{}}
	\toprule
	Used images              & Distance \\ \midrule
	All images (117-463)     & 0.291    \\
	Every 10th image (11-46) & 0.294    \\
	Every 20th image (5-23)  & 0.297    \\
	Every 50th image (2-9)   & 0.325    \\
	1 image                  & 0.699    \\ \bottomrule
\end{tabular}
\caption{This table shows the average distance if only a subsection of the training images are used.}
\label{tbl:fad-nth}
\end{table}

If the improvement best seen in Fig.~\ref{fig:dist-celeba-boxplot} is due to having access to different angles of the face, we would expect a similar improvement if we switch from using dozens of images to just a few picturing different angles.
Therefore, instead of using the template images in sequence, a greedy search on every iteration should result in the best embedding for each step.
On every step, we create the new average embedding for all remaining images of the person, calculate the new distance to the test images and select the one which minimizes this distance.
Table~\ref{tbl:fad-greedy} shows, that after using just the 3 best images, accuracy already improved significantly and there is little room for improvement (0.315 for the 3 best images vs 0.291 if all images are used).
After manually inspecting the top images for each person, in 82 \% of the cases the first 3 images are one frontal image, and two profile images from each side. 
Further work could add convergence criteria to automatically select the best amount of images.

\begin{table}[]
\centering
\begin{tabular}{@{}l|lll@{}}
	After n-th best images 	& 1 	& 2 	& 3 \\ \midrule
	Avg. distance 			& 0.571 & 0.370 & 0.315 \\
\end{tabular}
\caption{This table shows the average distance if images have been selected greedily.}
\label{tbl:fad-greedy}
\end{table}

\section{Related work}

Chowdhury \etal~\cite{chowdhury2016one} proposed an interesting change: Instead of using the mean-weighting of features, they propose to use the maximum instead.
This should reduce the overfit on dominant angles and generalize better~\cite{chowdhury2016one}. 
However, this could not be replicated with this dataset, as the \textit{minimum} and \textit{maximum} settings perform significantly worse than the baseline (Table~\ref{tbl:main-result}).
One potential cause for this bad performance is that outliers have too much impact on the final template.
Therefore, we created another template by using $\{25,75\}th$ quantile of each dimension of the embedding, which scores significantly better than both the \textit{minimum} and \textit{maximum} setting, but not as well as the \textit{average} aggregation strategy.

Rao \etal~\cite{rao2017attention} created a pipeline with a similar goal.
Instead of aggregating the embeddings into a single template, they created a neural network which receives raw images as input.
As the networks have full access to the whole image (instead of an embedding only), this approach offers the possibility of higher accuracy on the drastic expense of runtime-performance and is thus not really suitable for embedded systems.

Furthermore, in the last years, a lot of effort is spent on deciding how to intelligently weigh different dimensions of embeddings \cite{yang2017neural,rivero2021ordered,liu2019feature}.
Even though some of these approaches look promising, they are not ideal for embedded systems, as most of them use additional hardware-intense computations.
Therefore, this work does not favor any specific image over another.

Balsdon \etal\cite{balsdon2018improving} showed that accuracy of humans doing face identification significantly improves in a ``wisdom of crowd'' setting compared to individual's performance.
This could indicate, that a similar effect is demonstrable if a system combines embeddings not only from a single face recognition neural network, but from multiple different ones.
Therefore, further work could use the proposed method of combining embeddings of different neural networks, potentially using the same aggregation strategies as analyzed in the present paper.

\section{Conclusion}

In this work, we evaluated different aggregation strategies, leading to the conclusion that aggregating embeddings by taking the average of each dimension provides the highest improvement in accuracy while remaining compatible to state-of-the-art face recognition pipelines as already widely deployed in the field. We stress that this was one of the design goals of our work, and that our results indicate that such improvements can be directly applied to existing (embedded and distributed) systems with changes to only the enrollment and template computation processes, but not the live recognition pipelines.

Even though some previous work implicitly used this average aggregation strategy, there has been no evaluation about its effectiveness.
We base this proposal on an extensive evaluation of different aggregation strategies using both different public datasets and creating a new dataset, which is publicly available for research purposes.
After quantitatively analyzing the number of images used to generate templates, we find that it only plays a minor role, while different perspectives --- we refer to them as semantically different input --- significantly improve the performance of face recognition pipelines.
For an efficient, decentralized system, we propose to use just 3 images per template: one frontal image and one from each side.
These images may share the same setting, thus if there is a physical enrollment, these images can be taken live.
This increases both the correctness of the system itself (as there are fewer options to spoof the system) and the usability of the system (as the user does not have to provide larger sets of images or even video footage).

Future work could focus on automatically choosing the best images based on various convergence criteria, as well as further studying continuous learning approaches that update the aggregated template with new input as user faces change over time (e.g., with age or with seasons/clothing). We argue that the average aggregation strategy that results in best results in our study would lend itself optimally to dynamic updates; however, the associated security impact and added threat models by allowing templates to be updated outside a controlled enrollment setting will need to be considered carefully for each scenario.

\section*{\uppercase{Acknowledgements}}

This work has been carried out within the scope of Digidow, the Christian Doppler Laboratory for Private Digital Authentication in the Physical World and has partially been supported by the LIT Secure and Correct Systems Lab.
We gratefully acknowledge financial support by the Austrian Federal Ministry of Labour and Economy, the National Foundation for Research, Technology and Development, the Christian Doppler Research Association, 3 Banken IT GmbH, ekey biometric systems GmbH, Kepler Universit\"atsklinikum GmbH, NXP Semiconductors Austria GmbH \& Co KG, \"Osterreichische Staatsdruckerei GmbH, and the State of Upper Austria.

\bibliographystyle{apalike}
{\small
	\bibliography{bib}}

\begin{thebibliography}{}

\bibitem[Balsdon et~al., 2018]{balsdon2018improving}
Balsdon, T., Summersby, S., Kemp, R.~I., and White, D. (2018).
\newblock Improving face identification with specialist teams.
\newblock {\em Cognitive Research: Principles and Implications}, 3(1):1--13.

\bibitem[Chowdhury et~al., 2016]{chowdhury2016one}
Chowdhury, A.~R., Lin, T.-Y., Maji, S., and Learned-Miller, E. (2016).
\newblock One-to-many face recognition with bilinear cnns.
\newblock In {\em 2016 IEEE Winter Conference on Applications of Computer
  Vision (WACV)}, pages 1--9. IEEE.

\bibitem[Deng et~al., 2020]{deng2020retinaface}
Deng, J., Guo, J., Ververas, E., Kotsia, I., and Zafeiriou, S. (2020).
\newblock Retinaface: Single-shot multi-level face localisation in the wild.
\newblock In {\em Proceedings of the IEEE/CVF Conference on Computer Vision and
  Pattern Recognition}, pages 5203--5212.

\bibitem[Deng et~al., 2019]{deng2019arcface}
Deng, J., Guo, J., Xue, N., and Zafeiriou, S. (2019).
\newblock Arcface: Additive angular margin loss for deep face recognition.
\newblock In {\em Proceedings of the IEEE/CVF Conference on Computer Vision and
  Pattern Recognition}, pages 4690--4699.

\bibitem[Department, 2022]{datageneration}
Department, S.~R. (2022).
\newblock { Volume of data/information created, captured, copied, and consumed
  worldwide from 2010 to 2025 }.
\newblock
  \url{https://www.statista.com/statistics/871513/worldwide-data-created/}.
\newblock Accessed: 2022-02-15.

\bibitem[F{\'a}bi{\'a}n and Guly{\'a}s, 2020]{fabian2020anonymizing}
F{\'a}bi{\'a}n, I. and Guly{\'a}s, G.~G. (2020).
\newblock De-anonymizing facial recognition embeddings.
\newblock {\em Infocommunications Journal}, 12(2):50--56.

\bibitem[Findling and Mayrhofer, 2013]{findling2013towards}
Findling, R.~D. and Mayrhofer, R. (2013).
\newblock Towards pan shot face unlock: Using biometric face information from
  different perspectives to unlock mobile devices.
\newblock {\em International Journal of Pervasive Computing and
  Communications}.

\bibitem[Gebru et~al., 2021]{gebru2021datasheets}
Gebru, T., Morgenstern, J., Vecchione, B., Vaughan, J.~W., Wallach, H., Iii,
  H.~D., and Crawford, K. (2021).
\newblock Datasheets for datasets.
\newblock {\em Communications of the ACM}, 64(12):86--92.

\bibitem[Gong et~al., 2019]{gong2019video}
Gong, S., Shi, Y., Kalka, N.~D., and Jain, A.~K. (2019).
\newblock Video face recognition: Component-wise feature aggregation network
  (c-fan).
\newblock In {\em 2019 International Conference on Biometrics (ICB)}, pages
  1--8. IEEE.

\bibitem[{Government of India}, 2022]{aadhaar}
{Government of India} (2022).
\newblock {Unique Identification Authority of India}.
\newblock \url{https://uidai.gov.in/}.
\newblock Online; accessed 2021-10-27.

\bibitem[Hofer et~al., 2023]{bib:2023-hofer-aaiml}
Hofer, P., Roland, M., Mayrhofer, R., and Schwarz, P. (2023).
\newblock {Optimizing Distributed Face Recognition Systems through Efficient
  Aggregation of Facial Embeddings}.
\newblock {\em Advances in Artificial Intelligence and Machine Learning},
  3(1):693--711.

\bibitem[Huang et~al., 2008]{huang2008labeled}
Huang, G.~B., Mattar, M., Berg, T., and Learned-Miller, E. (2008).
\newblock Labeled faces in the wild: A database forstudying face recognition in
  unconstrained environments.
\newblock In {\em Workshop on faces in'Real-Life'Images: detection, alignment,
  and recognition}.

\bibitem[intersoft consulting, 2022]{gdpr}
intersoft consulting (2022).
\newblock { General Data Protection Regulation }.
\newblock \url{https://gdpr-info.eu/art-20-gdpr/}.
\newblock Accessed: 2022-08-16.

\bibitem[Liu et~al., 2019]{liu2019feature}
Liu, Z., Hu, H., Bai, J., Li, S., and Lian, S. (2019).
\newblock Feature aggregation network for video face recognition.
\newblock In {\em Proceedings of the IEEE/CVF International Conference on
  Computer Vision Workshops}.

\bibitem[Liu et~al., 2015]{liu2015deep}
Liu, Z., Luo, P., Wang, X., and Tang, X. (2015).
\newblock Deep learning face attributes in the wild.
\newblock In {\em Proceedings of the IEEE International Conference on Computer
  Vision}, pages 3730--3738.

\bibitem[Rao et~al., 2017]{rao2017attention}
Rao, Y., Lu, J., and Zhou, J. (2017).
\newblock Attention-aware deep reinforcement learning for video face
  recognition.
\newblock In {\em Proceedings of the IEEE international conference on computer
  vision}, pages 3931--3940.

\bibitem[Rivero-Hern{\'a}ndez et~al., 2021]{rivero2021ordered}
Rivero-Hern{\'a}ndez, J., Morales-Gonz{\'a}lez, A., Denis, L.~G., and
  M{\'e}ndez-V{\'a}zquez, H. (2021).
\newblock Ordered weighted aggregation networks for video face recognition.
\newblock {\em Pattern Recognition Letters}, 146:237--243.

\bibitem[Sauer, 2022]{moscow}
Sauer, P. (2022).
\newblock {Privacy fears as Moscow metro rolls out facial recognition pay
  system}.
\newblock
  \url{https://www.theguardian.com/world/2021/oct/15/privacy-fears-moscow-metro-rolls-out-facial-recognition-pay-system}.
\newblock Accessed: 2022-02-15.

\bibitem[Yang et~al., 2017]{yang2017neural}
Yang, J., Ren, P., Zhang, D., Chen, D., Wen, F., Li, H., and Hua, G. (2017).
\newblock Neural aggregation network for video face recognition.
\newblock In {\em Proceedings of the IEEE Conference on Computer Vision and
  Pattern Recognition}, pages 4362--4371.

\bibitem[Zheng et~al., 2020]{zheng2020automatic}
Zheng, J., Ranjan, R., Chen, C.-H., Chen, J.-C., Castillo, C.~D., and
  Chellappa, R. (2020).
\newblock An automatic system for unconstrained video-based face recognition.
\newblock {\em IEEE Transactions on Biometrics, Behavior, and Identity
  Science}, 2(3):194--209.

\end{thebibliography}

\end{document}